# Strictures of the female reproductive tract impose fierce competition to select for highly motile sperm

Meisam Zaferani[1], Gianpiero D. Palermo[2], Alireza Abbaspourrad[1,*]

[1] Department of Food Science and Technology, Cornell University, Ithaca, NY 14853, USA

[2] The Ronald O. Perelman and Claudia Cohen Center for Reproductive Medicine, Weill Cornell Medicine, New York, NY 10021, USA

*Corresponding author: alireza@cornell.edu, (607) 255-2923

**Abstract**

Investigating sperm locomotion in the presence of an external fluid flow and geometries simulating the female reproductive tract can lead to a better understanding of sperm motion during the fertilization process. In this study, using a microfluidic device featuring a stricture that simulates the biophysical properties of narrow junctions inside the female reproductive tract, we observed the gate-like role the stricture plays to prevent sperm featuring motility below a certain threshold from advancing towards the fertilization site. At the same time, all sperm slower than the threshold motility accumulate before the stricture and swim in a butterfly-shaped path between the channel walls which maintains the chance of penetrating the stricture and thus advancing towards the egg. Interestingly, the accumulation of sperm before the stricture occurs in a hierarchical manner so that sperm with higher velocities remain closer to each other and as the sperm velocity drops, they spread further apart.

**Keywords:**

**Sperm competition, Butterfly-shaped motion, Steering mechanisms, motility-based selection**

**Introduction**

In mammals, the number of sperm entering the female reproductive tract (~60–100 million) exceeds the number of available eggs (one egg during every ovulation) by far[1]. Accordingly, only a few sperm can fertilize the available eggs. Since motility is required for sperm to traverse the female genital tract[2], it has been thought that normal motility is one of the critical properties that determine the sperm's fertilization chances[3]. Consequently, motility-based competition must take place so that sperm with higher motility have a greater chance of fertilizing the egg[4].

In addition to motility, sperm require steering mechanisms to swim on the correct path towards the egg[5]. Chemotaxis[6] and thermotaxis[7] have been identified as steering mechanisms for marine invertebrate sperm, such as sea urchin[8–14]. However, their role in the guidance of mammalian sperm towards the egg is disputable[15]. In mammals, the fluid mechanical steering mechanisms of sperm include the tendency to follow rigid boundaries[16, 17] and swim upstream[17–20] (i.e., sperm rheotaxis). Swimming along rigid boundaries enables sperm to move parallel to the walls of the female reproductive tract, and the rheotactic behavior leads to their ability to swim opposite to the directional flow of secreted genital mucus[5, 22–24]. Since, the fluid mechanical steering mechanisms solely guide motile sperm, all non-motile sperm are carried away by the genital mucus flow while the healthy, motile sperm advance towards the fertilization site[24]. This tendency of the motile sperm to swim counterflow along the walls have been inspiring to design new microfluidic tool to hasten the process of sperm separation required for assisted reproductive technologies[25].

To date, the fluid mechanical steering mechanisms of mammalian sperm have been examined exclusively in straight swimming channels[18–20, 24]. However, the biophysical/fluid mechanical conditions of the female reproductive tract are more complex. In fact, the sperm swimming channel within the female reproductive tract does not have constant dimensions, but rather varies in width(26). Deviation in width of the swimming channel results in alteration of the flow magnitude, which consequently influences the sperm rheotactic and boundary swimming behavior. Therefore, investigation of sperm locomotion in a channel that mimics the biophysical aspects of the swimming channel *in vivo* will reveal the impact of the channel geometry on the sperm steering mechanisms(24). By the same token, the effect of the fluid mechanical properties of the female reproductive tract on motility-based competition among sperm cells can be revealed.

In this work, we examine sperm motion, including their fluid mechanical steering mechanisms, by solving sperm equations of motion inside a quasi-2D microfluidic design featuring variable width. Additionally, by experimentally observing sperm locomotion within the design, we show that strictures inside the sperm swimming channel play a gate-like role. That is, sperm slower than a threshold velocity cannot pass through the stricture, revealing the function of narrow junctions in the reproductive tract in selecting for highly motile sperm. Interestingly, sperm slower than the threshold velocity resist against the flow and accumulate before the stricture in a hierarchical manner in which motility-based competition becomes fiercer among highly motile sperm.

**Results and Discussion**

We designed and fabricated a quasi-2D microfluidic device (30 µm in depth) using conventional soft lithography that featured three eye-shaped compartments connected to each other by a progressive narrowing in width of the microchannel (Fig. 1a). The width of the channel in the narrowest section (i.e., the stricture) was 40 µm while the maximum width of each compartment was 300 µm. The angle of the stricture mouth was ~80°. The velocity field of the sperm medium (i.e., shear rate along the $\hat{z}$ and $\hat{y}$ directions) within each stricture was designed to be high enough to act as a barrier so that no sperm can pass through. The velocity field within the quasi-2D microfluidic channel was obtained by solving the conservation of momentum and mass equations with no-slip boundary conditions using finite element method simulations. The velocity field in an X-Y cut plane at a Z position corresponding to half the channel depth is demonstrated in Fig. 1(a), which shows the mean velocity field increases as the width of the channel decreases. In Fig. 1(b), the velocity profile of the fluid in four different cross sections is demonstrated using contour levels (X = 0, 25, 50 and 75 µm). According to the simulations, the maximum velocity field (125 µm/s) occurred in the stricture of the channel (X = 0) and decreased to as low as 20 µm/s at X = 300 µm. To use these numerical results for simulating the rheotactic behavior of the sperm, we extracted the shear rate of the fluid on the top surface of the chip in the $\hat{z}$ direction using $\gamma_z = \frac{\partial v}{\partial z}$, as shown in Fig. 1(c), in which $v$ is sperm medium velocity field. Moreover, to find the shear rate near the sidewalls in the $\hat{n}$ direction – i.e. unit vector normal to the sidewalls- , $\gamma_n = \frac{\partial v}{\partial n}$ was used.

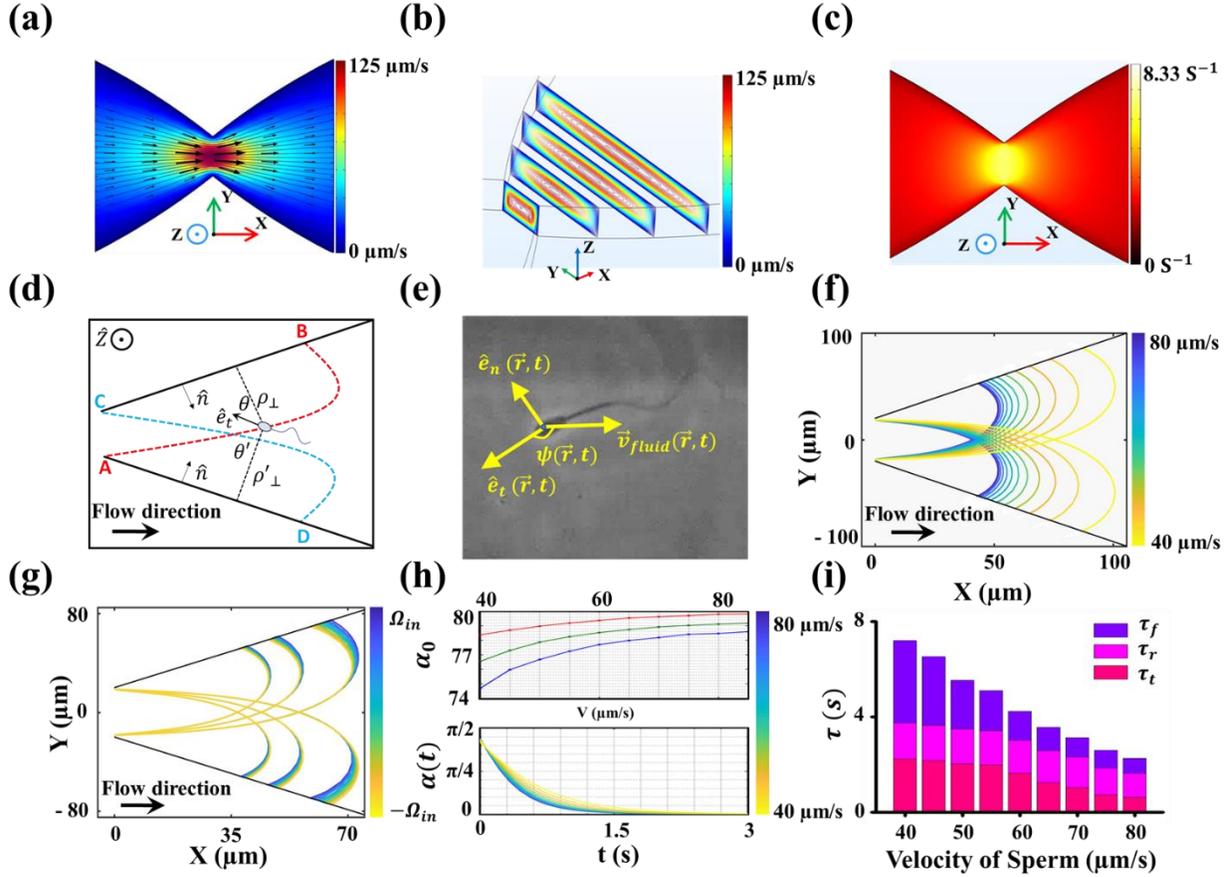

**Fig. 1 Simulation of sperm motion before the stricture.** (a) 2D velocity field of the sperm medium within the device at Z = 15 µm. (b) The velocity field of the medium demonstrated in YZ cut planes using contour levels. (c) The shear rate in proximity of the top surface of the channel. (d) Schematic of sperm butterfly-shaped motion, with depiction of all the variables. (e) Microscopic image of the sperm and the direction of flow. (f) The sperm path before the stricture for sperm with different velocities (40–80 µm/s). (g) The influence of $\Omega_{IN}$ on the sperm path. The value used as $\Omega_{IN}$ was experimentally measured as 0.12 ± 0.06 s$^{-1}$. (h) Top, the initial angle of the sperm with the sidewall at the contact point for $\Omega_{IN} = -\Omega_{max}, 0,$ and $\Omega_{max}$, illustrated with red, green, and blue respectively. Bottom, the time required for sperm to rotate upstream towards the stricture. (i) The total period (τ) required for sperm to depart from point A (C) and reach point C (A). The time elapsed in each mode is illustrated separately so that $\tau_f$, $\tau_r$, and $\tau_t$ correspond to the boundary, rotation, and transfer mode times.

To simulate the swimming path of the sperm, we assumed that the sperm location is influenced by its propulsive velocity, the velocity field of the medium, and the velocity components induced by the hydrodynamic interaction with the sidewalls. The velocity components induced by the sidewalls is described by the established far field approximation for a dipole pusher swimmer(27). The reflection of each image system on the other sidewall was neglected(23). Consequently, the sperm location could be modeled using the following equation:

$$\frac{d\vec{r}}{dt} = \vec{v}_{sperm} + \vec{v}_{fluid} + \frac{d\vec{\rho}_\perp}{dt} + \frac{d\vec{\rho'}_\perp}{dt} \quad [1]$$

in which $\vec{v}_{sperm}$ is the sperm propulsive velocity in the absence of sidewalls and fluid flow, and $\vec{v}_{fluid}$ is the sperm medium velocity field within the microfluidic channel at $Z = h - \delta$, where h is the channel height and the $\delta$ value reported for sperm is ~10 µm (Fig. S1). The two rightmost terms in Eq. 1 represent the drift velocity components induced by hydrodynamic interactions of the sperm with the sidewalls. $\rho_\perp$ and $\rho'_\perp$ are perpendicular distances of the sperm from the sidewalls, as shown in Fig. 1(d). Berke et al(28), reported that the hydrodynamic interaction terms are inversely correlated to the perpendicular distance of the sperm from the sidewalls, as described by:

$$\frac{d\vec{\rho}_\perp}{dt} = -\frac{3P}{64\pi\eta\rho_\perp^3}(1 - 3\cos^2\theta)\vec{\rho}_\perp \quad [2]$$

$$\frac{d\vec{\rho'}_\perp}{dt} = -\frac{3P}{64\pi\eta\rho'_\perp^3}(1 - 3\cos^2\theta')\vec{\rho'}_\perp \quad [3]$$

in which $P$ is the dipole strength of the sperm, $\eta$ is the viscosity of the sperm medium, and $\theta$ and $\theta'$ are the angles between the swimming direction of the sperm and the sidewalls. Based on equations (2) and (3), when $|\cos\theta| \leq 1/\sqrt{3}$, the sidewalls repel the sperm; otherwise, they attract it. To calculate the vertical distance of the sperm from the sidewalls, we found the microswimmer's minimum distance from the sidewalls at each point simulated in the channel. Generally, the influence of velocity components induced by hydrodynamic interactions on sperm motion is considered a secondary role and is small in comparison with the progressive velocity of the sperm and fluid flow of the medium(29). Moreover, the minimum distance between the sperm cell and the sidewall is 10 µm as explained in Fig. S1. Therefore, for $\rho_\perp$ or $\rho'_\perp$ < 20 µm, the drift velocity components induced by hydrodynamic interactions, assumed to remain constant and equal to those at $\rho_\perp$ (or $\rho'_\perp$) = 20 µm.

We considered the magnitude of the $\vec{v}_{sperm}$ as a constant value over time, while sperm energy loss was not considered in this paper. However, sperm swimming direction evolves with time and the sperm angular velocity is affected by its fluid mechanical response to an external fluid, intrinsic rotation, and response to the sidewalls (i.e. hydrodynamic interaction). Therefore, and based on the superposition principle, the angular velocity of the sperm at each point between the sidewalls can be described by:

$$\vec{\Omega}_{sperm}(\vec{r},t) = (-\Omega_{RH} - \Omega_{IN} - \Omega_{HI})\hat{z} \quad [4]$$

in which $\Omega_{RH}$ is the rheotactic angular velocity caused by the response of the sperm to the fluid flow. Tung et al.(22) described this rotation with Eq. 5:

$$\Omega_{RH} = \lambda \gamma_z \sin \psi(\vec{r},t) \quad [5]$$

in which $\lambda$ is a dimensionless constant related to the asymmetry in the sperm geometry, $\gamma_z$ is the shear rate in the vicinity of the top surface along the -$\hat{z}$ direction (Fig. 1(c)), and as shown in Fig.

1(e), $\psi(\vec{r}, t)$ is the angle between the direction of the sperm movement and the velocity field, which can be obtained by $\psi(\vec{r}, t) = \cos^{-1}(\frac{\vec{v}_{sperm} \cdot \vec{v}_{fluid}}{|\vec{v}_{sperm}| \cdot |\vec{v}_{fluid}|})$.

The intrinsic rotation ($\Omega_{IN}$) is the angular velocity created from asymmetry in the beating pattern of the sperm tail. The sperm flagellum does not feature a sine wave with a single frequency and phase. Rather, the mechanical wave produced by the sperm tail encompasses sine waves with different frequencies and initial phases. Consequently, the sperm tail beating pattern is asymmetric, which yields to an intrinsic rotation(15) (see the Intrinsic Angular Velocity section of the Supplementary Information). This rotation can be modeled by an intrinsic angular velocity that is described by Eq. 6:

$$\Omega_{IN} \propto v_{propulsion} \left(\frac{\xi_t}{\xi_n}\right) \frac{y_n}{L^2} \cos(\phi) \quad [6]$$

in which $\xi_t$ and $\xi_n$ are tangential and normal friction coefficients of the sperm medium, $y_n$ is the amplitude of the n$^{th}$ harmonic of the sperm tail, $L$ is the length of the sperm tail, and $\phi$ is the phase difference between the main sine wave and the n$^{th}$ harmonic. Since $\phi$ is an arbitrary parameter, $\Omega_{IN}$ can vary from $-\Omega_{max}$ to $\Omega_{max}$, which means its rotation can be either counterclockwise or clockwise. To experimentally measure the intrinsic angular velocity (Movie S1, Fig. S2 and S3), we extracted the trajectories of 28 sperm trapped in an area with no background fluid flow. Interestingly, the majority of the sperm were moving clockwise and the value we measured for those sperm was $\langle \Omega_{IN} \rangle = 0.12 \pm 0.06 \ s^{-1}$. Accordingly, the absolute value of the intrinsic angular velocity in all simulations was assumed to be less or equal to $\langle \Omega_{IN} \rangle$ ($|\Omega_{IN}| \leq \langle \Omega_{IN} \rangle = \Omega_{max}$).

The rotation caused by hydrodynamic interactions with the sidewalls, as reported by Berke et al(30)., can be described by Eq.7:

$$\Omega_{HI}(\vec{r},t) = -\frac{3P\cos\theta\sin\theta\,(3+\cos^2\theta)}{128\eta\,\rho_\perp^3} + \frac{3P\cos\theta'\sin\theta'\,(3+\cos^2\theta')}{128\eta\,\rho'^{\,3}_\perp} \quad [7]$$

For simplicity, the white zero-mean Gaussian noise in the sperm head rotation was also neglected. Finally, we can write the time-derivative of the sperm swimming direction $\hat{e}_t(\vec{r},t)$ as the outer product of this unit vector with the angular velocity vector of the sperm:

$$\dot{\hat{e}}_t(\vec{r},t) = \vec{\Omega}(\vec{r},t) \times \hat{e}_t(\vec{r},t) \quad [8]$$

For $\Omega_{IN} = 0$ and the initial sperm orientation parallel to the sidewall, the trajectory calculated for the sperm with different velocities is presented in Fig. 1(f). Since the shear rate within the stricture is too high for sperm to pass, the swimmer detaches from the sidewall at the vicinity of the stricture and is swept away by the flow until it reaches the other sidewall. The sperm trajectory depicted in Fig. 1(f) shows that the location of the initial contact point of the sperm and this sidewall is not linearly related to the sperm velocity. That is, as the sperm velocity declines, the sperm are increasingly carried away by the fluid flow. To demonstrate the impact of intrinsic rotation on the sperm locomotion, the trajectories of the swimmers with velocities of 40, 50, and 60 µm/s are also depicted in Fig. 1(g) for $|\Omega_{IN}| \leq \Omega_{max}$. As can be seen, the effect of intrinsic rotation is more substantial on slower sperm (40 µm/s) whereas the increase/decrease in location of the initial contact point caused by $\Omega_{IN}$ is smaller for faster sperm (60 µm/s).

Upon arrival to the opposing sidewall, the shear rate along $\hat{n}$ (i.e. $\gamma_n$) starts rotating the sperm upstream. Depending on the angle between the sperm orientation and the sidewall at the contact point (Fig. 1(h), top), and the sperm velocity, which determines the location of the contact point, the rotation time will vary (Fig. 1(h), bottom). In Fig. 1(h) top, the initial angle of the sperm with the sidewall is depicted for different velocities, with the red, green, and blue curves corresponding to $\Omega_{IN} = -\Omega_{max}, 0$, and $\Omega_{max}$, respectively. The evolution of the angle between the sperm orientation and the sidewall at the contact point is determined by Eq. 5, in which $\gamma_z$ is replaced with $\gamma_n$. Therefore, the time required for the sperm to reorient itself upstream is the time required to decrease its angle with the sidewall from $\alpha_0$ to $\delta/L \sim \frac{\pi}{20}$ (Fig. S1). After the upstream orientation, the sperm starts following the sidewall. This boundary movement is simply determined by the shear rate along the $\hat{n}$ direction multiplied by $\delta$ subtracted from its propulsive velocity. Finally, the total time required for sperm to return to its initial X coordinate ($A \rightarrow C$), $\tau$ is shown in Fig. 1(i) for different sperm velocities. $\tau$ includes the time required for sperm to transfer from one sidewall to the other ($\tau_t$), rotate upstream at the contact point B/D ($\tau_r$), and follow the boundary in the stricture direction ($\tau_f$). Clearly, the time required for sperm to return to the initial X and Y coordinates is $2 \times \tau$.

Based on the simulation, sperm movement before the stricture –i.e. hydrodynamic barrier- is comprised of three different modes: (1) transfer mode, in which the sperm detaches from the sidewall, becomes swept back by the flow and reaches the opposing sidewall; (2) rotation mode, which involves sperm rotation around its head (i.e., the pivot) at the contact point; and (3) boundary swimming mode. When sperm motion begins at point A, its initial orientation is parallel to the sidewall, as can be seen in Fig. 1(d). The high shear rate at the mouth of the stricture causes the sperm to be swept back to point B on the opposing sidewall (i.e., transfer mode), at which point it stops moving perpendicular to the BC sidewall and starts rotating counterclockwise (i.e., rotation mode). By rotating near the wall, the sperm orients its direction parallel to the BC sidewall and begins moving along it in the direction of the stricture (i.e., boundary swimming mode). Upon arrival at point C, it detaches from the sidewall (similar to point A) due to the high shear rate of the structure and begins swimming towards the AD sidewall again. This periodic motion takes on a butterfly-shaped path ($A \to B \to C \to D \to A$) and continues until the sperm has no more energy to swim.

For a given angle between the two sidewalls ($\beta$), only sperm with velocities in a specific range can move in butterfly-shaped paths (Fig. S4). The upper limit of this range is determined by the shear rate in the stricture. The lower limit of the range, however, is determined by two conditions: (1) proximity to the stricture; and (2) return ability conditions. The proximity condition determines the velocity of all sperm that can become proximate to the stricture while the return ability condition means that at point B and D, the shear rate is adequate to reorient sperm towards upstream. Since the shear rate outside the stricture decreases, depending on the angle, sperm can get close to the stricture. By defining the proximity zone as $x < 5$ μm (the average size of the bull sperm head), for $\beta \sim 80°$, all the sperm with velocities between ~30–80 μm/s were able to become proximate to the stricture. By considering the return ability condition, we observed that among the sperm with velocities in this range, the fluid flow can reorient only sperm with velocities higher than 40 μm/s at point B (and D) in Fig. 1(d). In fact, the angle between the swimming direction and the sidewall at point B (and D) for sperm with motilities slower than 40 μm/s becomes greater than 90° and the shear rate in these points is inadequate to reorient the sperm upstream. Consequently, sperm slower than 40μm/s follow the boundary in the downstream direction.

Given the similarity between this stricture to the junctions of the sperm's path towards the site of fertilization, and the direction of the fluid flow, which simulates the mucus outflux within the tract, the final goal of the sperm in this situation is to pass through the stricture and advance towards the site of fertilization, or at least to maintain its location nearby the stricture. Since it is known that no sperm with velocities in the range of 40–80 µm/s can pass through the stricture, we defined the ability of the sperm to remain close to the stricture mouth as a competition index (CI). Since, the main path of the sperm towards the fertilization site is in the $-\hat{x}$ direction, we projected the sperm quasi-2D periodic motion onto the $x$ axis, as can be seen in the schematic of sperm motion in Fig. 2(a). Given the total period of the sperm motion ($T = \tau$), and neglecting the translational diffusivity of the sperm(31), the probability of the sperm to be closer to the stricture than at $x = a$ can be defined as:

$$CI = F_X(x) = pr\{X \le x = a\} = \int_0^a dx P(x,t) = \int_0^{T_a} \frac{dt}{T} + \int_{T'_a}^{T} \frac{dt}{T} = \frac{T_a + T - T'_a}{T} \quad [9]$$

in which $T_a$ and $T'_a$ are the times at which $x = a$ (Fokker-Planck equation in the Supplementary Information). We calculated the CI for different values of $a$ as a function of sperm velocity, the results of which are shown in Fig. 2(b).

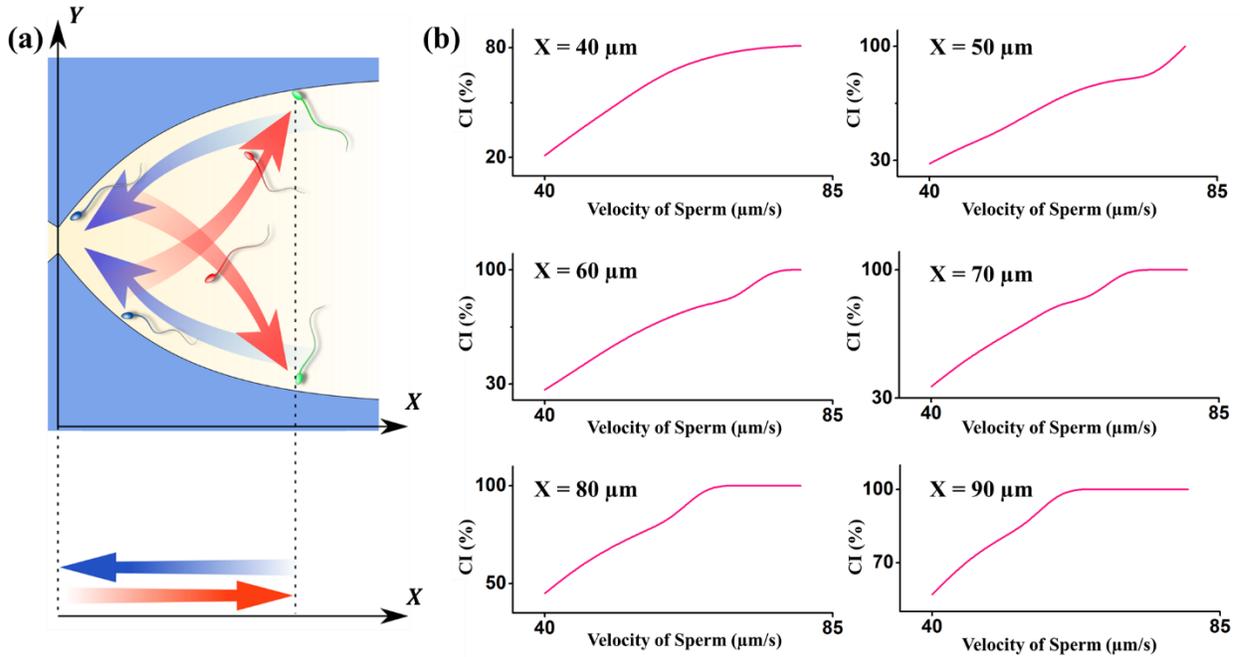

**Fig. 2 Sperm motion modes and the competition index.** (a) The schematic of sperm motion in different modes, including transfer, rotation, and boundary swimming modes, illustrated in red, green, and blue, respectively. The sperm projection in the X direction demonstrates a periodic motion, on which we based the competition index. (b) The competition index for sperm slower than a particular velocity drops depending on the value of a.

We also experimentally observed the butterfly-shaped motion described by the simulation in our microfluidic device. The butterfly-shaped motion of a bull sperm with a velocity of 54 µm/s is demonstrated in Fig. 3(a) and (b). Fig. 3(a) is a combined image of the sperm location at 23 different frames and Fig. 3(b) is the corresponding schematic of the sperm swimming pattern based on Fig. 3(a) for better visualization. To extract the sperm trajectories, we acquired videos of the device, as shown in Movie S2 in the Supplementary Information, and tracked the sperm movement over the elapsed time (i.e., 3–20 periods). Using MATLAB R2017a, we tracked 44, 35, and 51 sperm heads displaying different motilities to elucidate the trajectories of each microswimmer in three different sperm samples. Fig. 3(c) displays the motion of a single sperm at two different periods, in which the shape of the swimming path remained relatively constant over the elapsed time ($t_{blue} = 0 - 6.24$ s, $t_{red} = 38.14 - 44.11$ s). The motion of the sperm and its butterfly-shaped path, as can be seen in Fig. 2(c), is similar to the results obtained by simulations presented in Fig. 1(f).

To confirm the results obtained from the simulations, we experimentally measured the distance between the sperm-sidewall contact points (B and D) and detachment points (A and C), i.e., the "withdrawal distance," for sperm with different velocities. As shown in Fig. 3(d), the increasing velocity of the sperm led to a decay in the withdrawal distance (W). According to simulation- and experimental-based results, the decay in the withdrawal distance is exponentially correlated to the sperm velocity. That is, the difference in the withdrawal distances of two sperm is not only related to difference in the velocity of those swimmers—the velocity of the sperm plays a determinative role as well. For instance, based on experimental data, the mean withdrawal distances measured for sperm with velocities of ~75 µm/s and ~85 µm/s were 39.1 and 28.2 µm, respectively. However, for two slower sperm (~55 µm/s and ~65 µm/s) with the same difference in velocity, the corresponding withdrawal distances were 74.6 µm and 51.3 µm. This dependency of the sperm swimming path with the velocity of the sperm suggests that swimmers with higher velocities move closer to each other, and their corresponding CIs are closer in comparison than slower sperm.

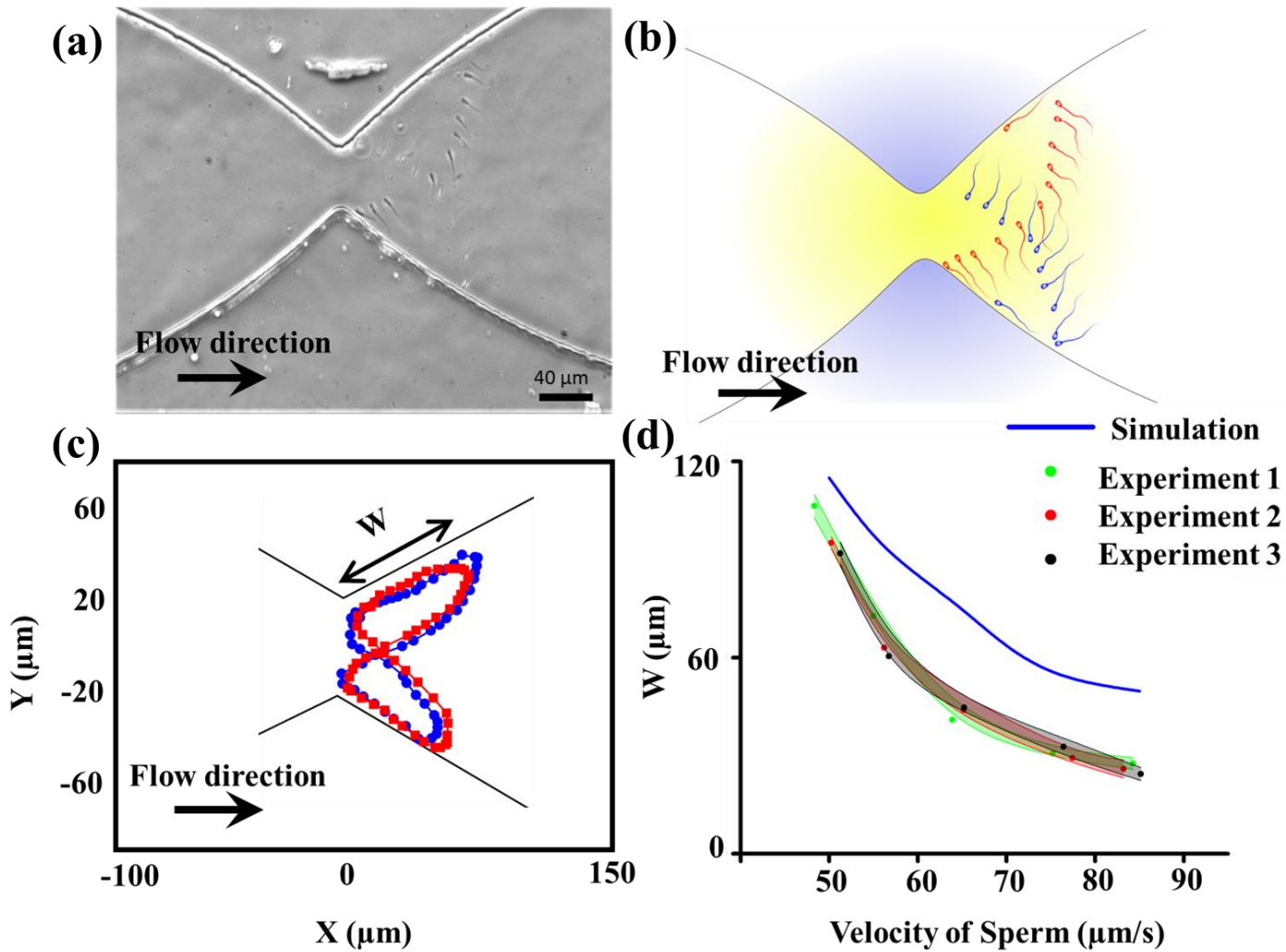

**Fig. 3 Observation of the butterfly-shaped swimming path and withdrawal distance of sperm before a stricture**. (a) The butterfly-shaped path extracted for a sperm with velocity of 57 µm/s. (b) The schematic of the butterfly-shaped path based on the experimental results obtained in part (a) for better visualization. (c) The trajectory of the sperm during two different periods to illustrate the consistency of the butterfly-shaped path over time. W is the withdrawal distance of the sperm. (d) Experimental values of the withdrawal distance extracted from 120 sperm with different velocities from three different samples in comparison with values expected by simulations.

In addition to withdrawal distance, we also measured the elapsed time taken for sperm to move in the transfer, rotation, and boundary swimming modes. The images acquired from sperm at different time frames and modes are presented in Fig. 4. In the pictures shown in Fig. 4(a), 5 sperm are moving in the transfer mode, in which three of them (colored blue) are departing the upper sidewall and moving towards the bottom sidewall. Likewise, the remaining two sperm (yellow-colored) are departing the bottom sidewall towards the upper sidewall. In Fig. 4(b), a moderately motile sperm with a velocity of ~59 µm/s can be seen beginning to rotate upstream due to the shear rate along the normal direction of the sidewall. Later on, this sperm swims along the boundary of the sidewall until it reaches the stricture, as can be seen in Fig. 4(c). These experimental observations of sperm movement thus help confirm our simulated-derivations of the three different sperm swimming modes.

The corresponding elapsed times of each mode ($\tau_t$, $\tau_r$, $\tau_f$) are presented in Fig. 5(a). Since the rotation and boundary swimming times somewhat overlap, especially for highly motile sperm, we combined the amount of time required for each of these modes into a single measurement ($\tau_r + \tau_f$). Using these times, we measured the CI from experimental data of sperm featuring different velocities for different values of $a$ (Fig. 5(b)). At $a = 40\ \mu m$, the CI measured for sperm with velocities higher than 78 µm/s was close to 100%, which means these sperm were always closer than 40 µm to the stricture. For velocities below 78 µm/s, the CI decays, with the CI for the slowest sperm ($v = 48$ µm/s) being 14.3%. In the case of $a = 50\ \mu m$, the range of velocity for sperm with CI of 100% expands and sperm faster than 72 µm/s are always closer than 50 µm to the stricture.

The similarity between human and bovine sperm in terms of the shape and swimming mechanism suggests that the motion of the human sperm before the stricture is like that of bovine sperm. We also experimentally observed human sperm motion before the stricture (Movie S3), and as was expected, the butterfly-shaped swimming path was seen in human sperm as well (Fig. S5). Moreover, the human sperm motion in the transfer, rotation and boundary swimming modes is also demonstrated in Fig. S6, which confirms the similarity between human and bull sperm locomotion strategies.

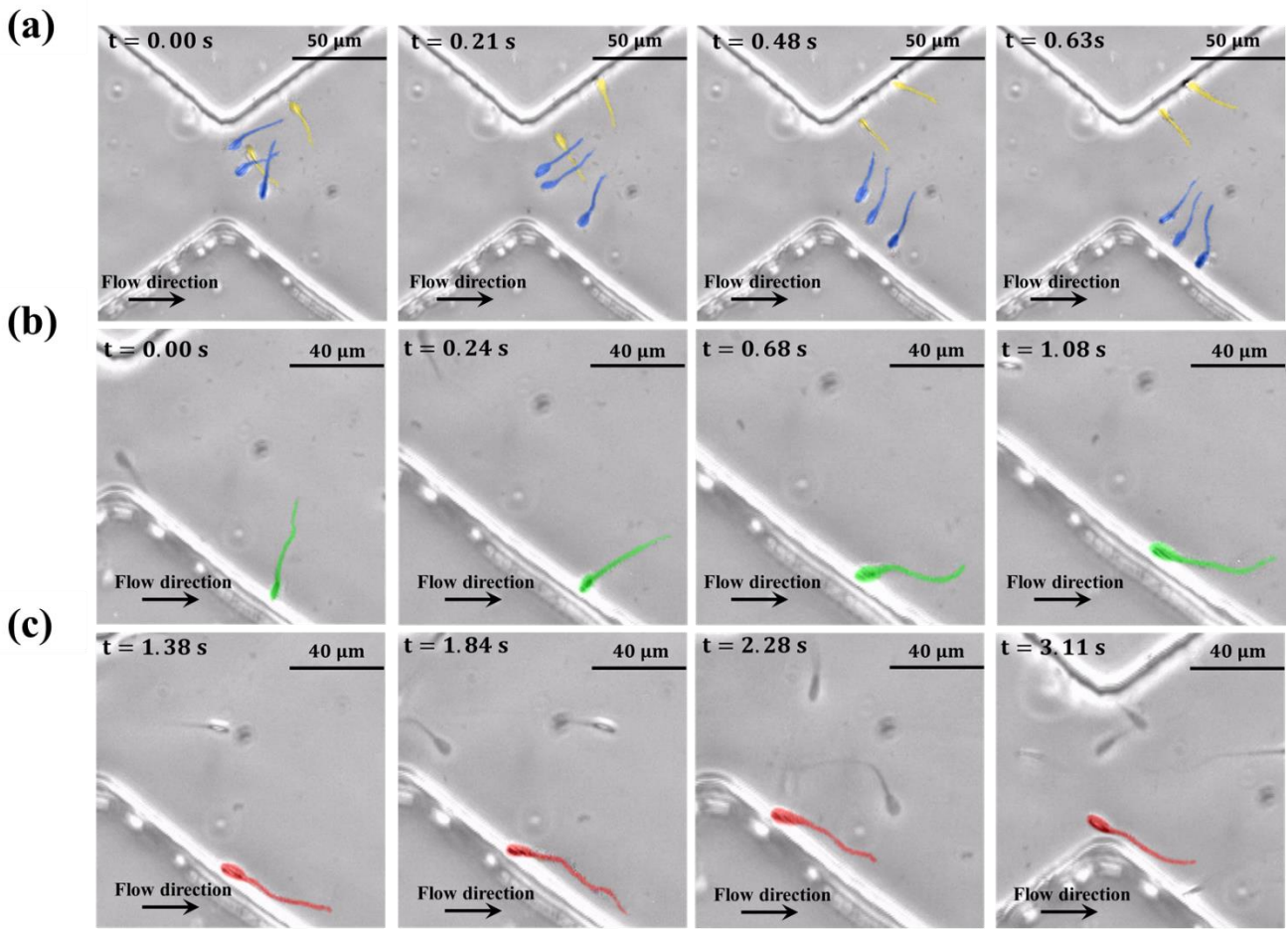

**Fig. 4 Experimentally measured times of the different swimming modes (i.e., transfer, rotation, and boundary swimming)**. (a) Images of sperm (blue and yellow) moving in transfer mode. (b) A single sperm (green) in rotation mode, reorienting upstream. (c) A sperm (red) in the boundary swimming mode begins following the sidewall.

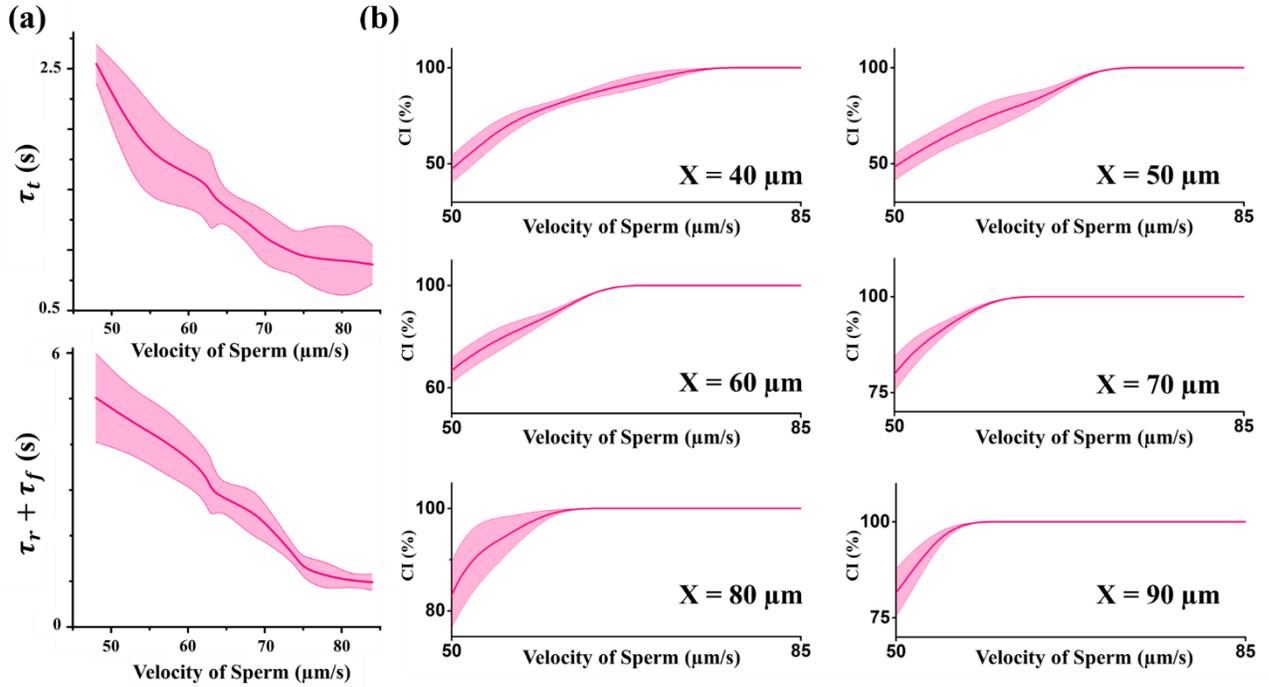

**Fig. 5 Experimental CI calculated for sperm**. (a) The times required for sperm to transfer ($\tau_t$), rotate ($\tau_r$) and follow the boundary ($\tau_f$) were experimentally measured for 120 sperm. Since the rotation and the boundary swimming times overlapped, their sum ($\tau_r + \tau_f$) was reported and measured. (b) For a given a, as the velocity of the sperm decreases, its likelihood to maintain its X coordinate closer than a decays.

## Accumulation of sperm near the stricture

In agreement with previous studies done in the absence of fluid flow(32, 33), the butterfly-shaped motion (due to the sperms' ability/tendency to swim counter to the flow and parallel to the sidewalls) also leads to the accumulation of sperm near the stricture, which could be interpreted as a mechanism used by the sperm to resist against the fluid flow. In fact, despite dead and non-motile sperm being carried away by the flow, motile sperm maintained their proximity to the stricture and thus their likelihood to pass through it is high. To observe this accumulation phenomenon, we observed the microfluidic device using low-magnification phase contrast microscopy(34) (Movie S4). To assess the abundancy of the sperm we took advantage of the twinkling effect observed in the motile bull sperm due to the paddle-shaped head of the microswimmers, in which the side of the sperm flashes bright under the imaging conditions, while the head's top and bottom face appear dark (Fig. 6(a)). Zone A of Fig. 6(b), which describes the area of the device that includes the stricture, features more twinkling, whereas zone B, which includes the wider region of the channel, twinkles less, as can be seen in Movie S4. Based on this evidence, we can conclude that more sperm accumulate near the stricture. To validate this twinkling effect-based result, we manually counted the total number of motile sperm swimming in zones A and B at high-magnification, the results of which are shown in Fig. 6(c) for three samples. With this method as well, we observed that the number of sperm that accumulate before the stricture is greater than that of zone B. This accumulation before the stricture demonstrates sperm resistivity against the flow, which leads to persistence upon advancing towards the egg, thus maintaining the chance of fertilization.

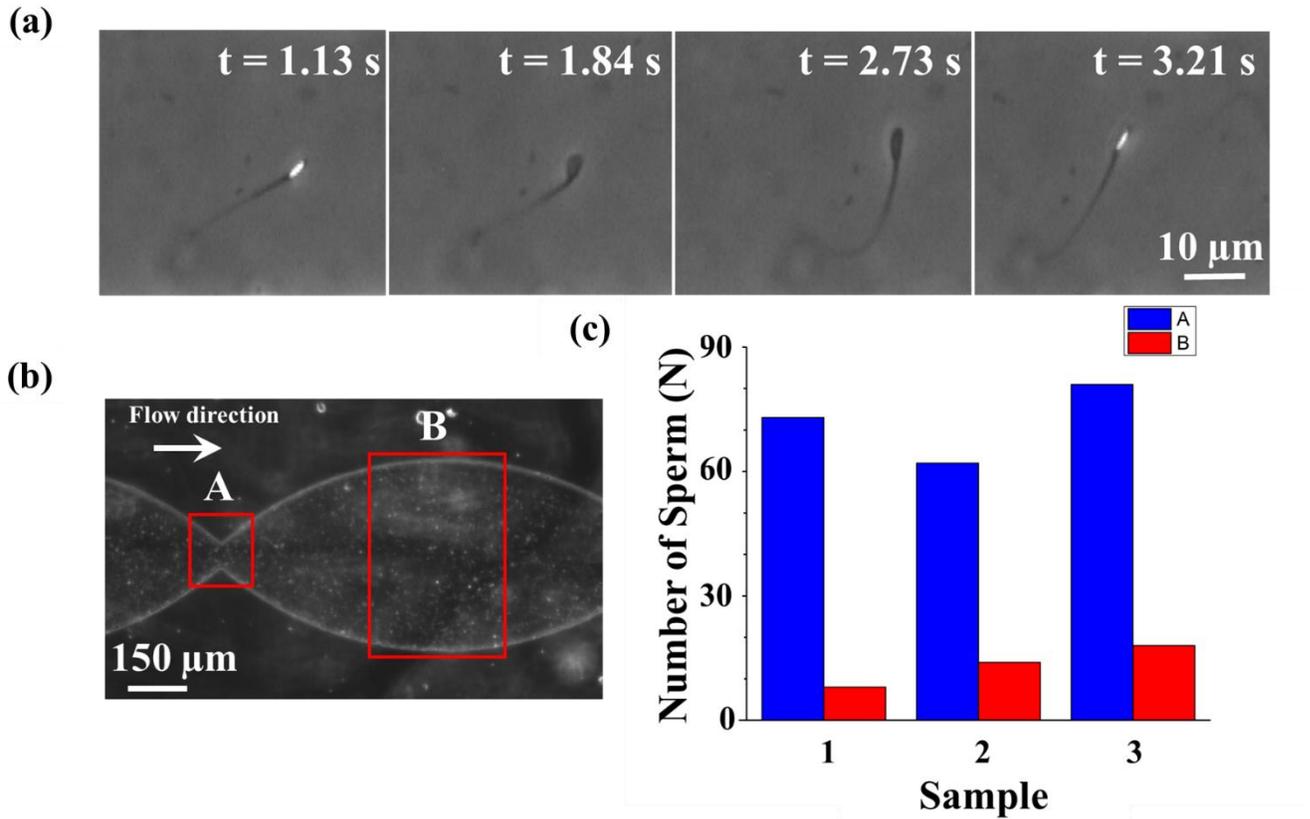

**Fig. 6 Accumulation of sperm before the stricture and the twinkling effect.** (a) Phase contrast microscopy leads to twinkling of the bull sperm. (b) Low magnification image of our device with a concentrated sample injected. Zones A and B are indicated in the image. (c) The number of live sperm in each zone for three different samples were counted manually to confirm the accumulation of the sperm.

**Gate-like role of the stricture**

The height of the hydro-mechanical barrier (i.e., the shear rate within the stricture) determines the threshold motility that sperm must possess to overcome and pass through the stricture. To experimentally observe the gate-like role of the stricture, we decreased the sperm medium injection flow rate and consequently the shear rate within the stricture to 7.98 s$^{-1}$. As a result, the sperm with the highest motility ($v = 84.2$ µm/s, sperm number 1) could resist against the flow within the stricture, as can be seen in Fig. 7(a) and Movie S5. Meanwhile, all the sperm with lower velocities (sperm numbers 2–6) maintained their location before the barrier by periodically moving between the sidewalls. In these conditions, sperm number 1 is almost static in the observer frame in the $\hat{x}$ direction, as neither the shear rate of the flow nor the sperm's motility can overcome the other. By further decreasing the shear rate of the stricture to 7.16 s$^{-1}$, eventually sperm number 1 can overcome the barrier and advance towards the next compartment, as can be seen in Fig. 7(b) and Movie S5. Meanwhile, all other sperm with lower velocities accumulate before the barrier in a hierarchical manner, i.e., sperm with higher velocities remain closer to the stricture and slower sperm are swept further back by the flow. The velocity of the sperm that passes the stricture -i.e. threshold velocity – is measured for different shear rates within the stricture and demonstrated in Fig. S7. The similarity between the geometry of the stricture and the junctions within the female reproductive tract suggests that the gate-like selective behavior of this microfluidic stricture can mimic the role of the junctions in the fertility process(2, 35).

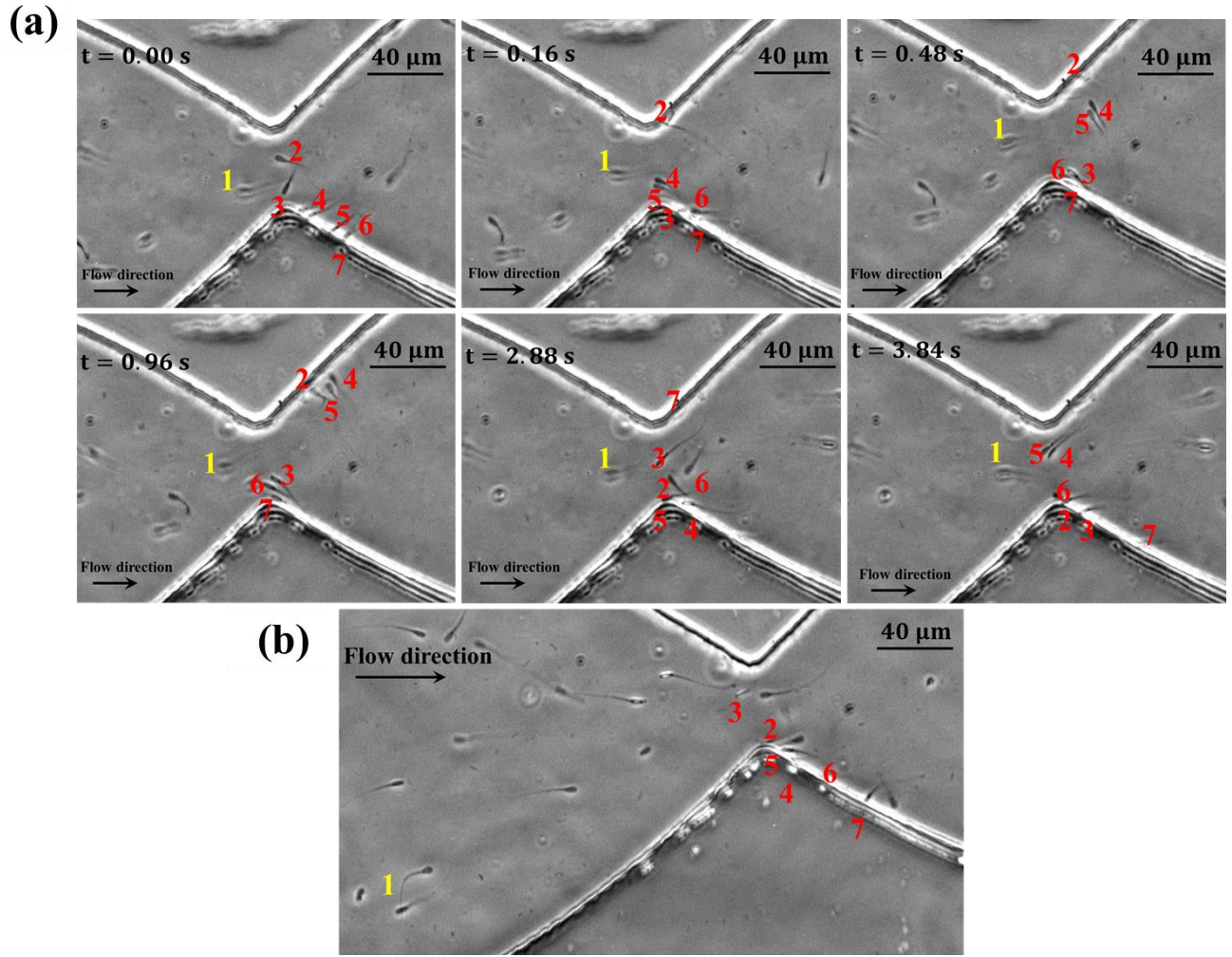

**Fig. 7 Gate-like role of the stricture.** (a) Sperm number one is able to resist against the shear rate in the stricture, and therefore it has approximately no movement in the -X direction. The other sperm (numbers 2-7) move in butterfly-shaped paths, but they cannot pass through the junction. Hierarchical swimming is discernible and sperm with higher velocity are closer to the stricture and to each other. (b) A small decrease (7.98 to 7.16 s$^{-1}$) in the injection flow rate led to sperm number 1 advancing and entering the adjacent compartment. Meanwhile, the slower sperm continue to swim on the butterfly-shaped path before the stricture.

**Conclusion**

The sperm response to fluid flow and their inclination to follow solid boundaries leads to sperm accumulation before the opening of the stricture inside a microfluidic design. The accumulated sperm featured a hierarchy, in which swimmers with higher motilities were closer to the stricture mouth, while slower sperm remained further apart away from the stricture. Using numerical simulations and experimental observations, we could quantitatively show that this hierarchical structure imposes competition among the sperm, with the fiercest occurring among highly motile microswimmers in comparison with the slower sperm.

Moreover, depending on the shear rate within the stricture, sperm with velocities higher than a threshold value can pass through the stricture whereas sperm slower than the threshold accumulate before the stricture. This gate-lake behavior of the stricture suggests a motility-based selection mechanism that may be used by the female reproductive tract to select for sperm with the highest motility. Since the flow rate of human genital mucus varies over time, this gate-like behavior shows that sperm location is maintained near the stricture till the shear rate within the stricture decreases; thus, the chance of the sperm to pass through the junction and advance towards the fertilization site is maximized. This investigation demonstrates that the geometry of the female reproductive tract plays a crucial role in motility-based sperm selection and competition so that highly motile sperm are most likely to pass through the fluid mechanical barriers and fertilize the egg.

**Materials and Methods**

**Bull and human sperm samples**

All the experiments were performed with four bull sperm samples frozen in 250 µL straws that were purchased from Genex Cooperative (Ithaca, NY). Semen from two of the bulls was frozen in a milk-based extender and semen from the other two bulls was frozen in an egg-yolk based extender at a concentration of 100 million sperm/mL. Frozen straws were thawed in a 37 °C water bath. Then the live sperm were separated from the dead sperm using a density gradient method(24). The separated sample was then diluted 1:3 using TALP (Tyrode's albumin lactate pyruvate) medium. The viscosity of the bull sperm sample after the dilution was 0.87 mPa·s at T = 37 ºC.

Fresh human sperm samples were generously provided by Weill Cornell Medical School. The original concentration of the human sperm sample was 46 million sperm/mL. All experiments carried out on human samples diluted 1:3 with TALP medium at T = 37 ºC. The viscosity of the human sample after the dilution was 0.94 mPa·s.

TALP recipe: NaCl (110 mM), KCl (2.68 mM), $NaH_2PO_4$ (0.36mM), $NaHCO_3$ (25 mM), $MgCl_2$ (0.49 mM), $CaCl_2$ (2.4 mM), HEPES buffer (25 mM), Glucose (5.56 mM), Pyruvic acid (1.0 mM), Penicillin-G (0.006% or 3mg/500 mL), BSA (20 mg/mL).

**Device fabrication and injection systems**

We used conventional soft lithography to fabricate the microfluidic device out of polydimethylsiloxane(36). Syringe pumps (Chemyx Fusion 200) were used to control the flow rate of the sperm medium at different injection rates of 0.6, 1.2, and 1.8 mL/h.

**Image and video acquisition**

Images and videos were acquired at 25 frames per second using phase contrast microscopy with a 10x objective and a digital Neo CMOS camera. During the experiments, the microfluidic chip was kept on a heated microscope stage (Carl Zeiss, at 37 ºC). The average path velocity of the sperm was determined using ImageJ (Version 1.51j8) and MATLAB (Version R2017a) software by measuring the average distance between the center of the sperm head in each frame divided by the time elapsed. This quantity is known and reported as VAP (average path velocity) in computer-assisted sperm analysis (C.A.S.A) systems.

**Simulation software**

The layout of the microfluidic device was imported into COMSOL MULTIPHYSICS (Version 5.2) simulation software. Using the laminar fluid module in stationary mode, we solved the Navier-Stokes (Eq. 10) and conservation of mass (Eq. 11) equations with a no-slip boundary condition at the sidewalls(37):

$$\rho(v.\nabla v) = -\nabla p + \nabla.\mu(\nabla v + (\nabla v)^T) \quad [10]$$

$$\nabla.v = 0 \quad [11]$$

in which v denotes the velocity field, ρ is the density of the sperm medium, p is pressure, and μ is the dynamic viscosity. To numerically solve the sperm equations of motion, MATLAB (Version R2017a) and an explicit Runge-Kutta method (i.e., the Dormand-Prince pair(38)) was used.

**Acknowledgements**

The authors would like to thank Dr. Soon Hon Cheong for providing the frozen bull semen samples and phase contrast microscopy. We also would like to thank Mary Godec for proofreading the paper and useful discussions, Farhad Javi for his intellectual inputs and assistance with the illustrations, and Philip Xie, Derek Keating and Alessandra Parrella for providing the human sperm sample. This work was performed in part at the Cornell NanoScale

Facility, a member of the National Nanotechnology Coordinated Infrastructure (NNCI), which is supported by the National Science Foundation (Grant ECCS-1542081).

# Strictures of the female reproductive tract impose fierce competition to select for highly motile sperm

Meisam Zaferani[1], Gianpiero D. Palermo[2], Alireza Abbaspourrad[1,*]

[1] Department of Food Science and Technology, Cornell University, Ithaca, NY 14853, USA

[2] The Ronald O. Perelman and Claudia Cohen Center for Reproductive Medicine, Weill Cornell Medicine, New York, NY 10021, USA

*Corresponding author: alireza@cornell.edu, (607) 255-2923

## Supplementary Information

**Delta (δ) value**

Sperm flagellum movement is not uniform. Instead, it has greater oscillation amplitude at the end of its tail in comparison to the head. This difference in oscillation amplitude at the end of the tail and the head leads to the tilted movement of the sperm in the withdrawal distance, and subsequently the tail experiences higher flow rate in comparison with the head (as the head is very close to the boundary). As a result, the shear rate close to each wall (sidewalls and top surface) plays the main role in sperm rheotaxis and the boundary swimming movement (Eq. 5). δ is a geometrical value used to describe this tilted orientation[6], as shown in the Fig. S1. According to the δ value reported for sperm, the final orientation of the sperm during the boundary swimming movement is $\delta/L \sim \frac{\pi}{20}$, in which L is the sperm length. The tilted orientation of the sperm therefore makes it susceptible to the flow of the sperm medium.

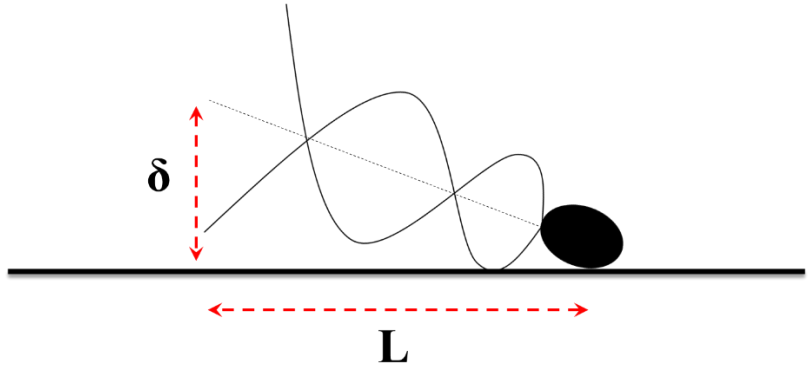

**Fig. S1 Sperm tilted orientation in the boundary swimming mode.**

**Intrinsic angular velocity of sperm**

**Intrinsic angular velocity**

To derive the angular velocity of a sperm imposed by its periodic flagellum movement, we assumed that flagellum shape over time is a sum of simple sine waves[1,2] with the amplitudes of $y_i$, temporal frequencies of $\omega_i$, and phases of $\phi_i$. For simplicity, we assumed that all the sine waves were moving with an identical wave number $k$:

$$y(x,t) = \sum y_i \sin(kx - \omega_i t + \phi_i) \qquad [S1]$$

Therefore, the vertical component of the velocity of a segment of the flagellum is the time-derivative of $y(x,t)$, whereas the horizontal component of the velocity is zero (Eq. S2). Also, to apply resistive force theory and derive the expression for sperm movement, its velocity was decomposed into tangential and normal components.

$$\vec{V} = \left(0, \frac{\partial y}{\partial t}\right) \qquad [S2]$$

To decompose this velocity into tangential and normal components, the tangent and normal unit vectors (Eq. S3 & 4) must be plugged into Eq. S5, which is a standard notion of force-velocity relation in established resistive force theory[2,3], in which $\xi_n$ and $\xi_t$ are anisotropic friction coefficients of the normal and tangential directions, respectively.

$$\hat{e}_t = \frac{1}{\sqrt{1+\left(\frac{\partial y}{\partial x}\right)^2}}\left(1,\frac{\partial y}{\partial x}\right) \approx \left(1-\frac{1}{2}\left(\frac{\partial y}{\partial x}\right)^2\right)\left(1,\frac{\partial y}{\partial x}\right) \qquad [S3]$$

$$\hat{e}_n = \frac{1}{\sqrt{1+\left(\frac{\partial y}{\partial x}\right)^2}}\left(-\frac{\partial y}{\partial x},1\right) \approx \left(1-\frac{1}{2}\left(\frac{\partial y}{\partial x}\right)^2\right)\left(-\frac{\partial y}{\partial x},1\right) \qquad [S4]$$

$$\vec{f} = -\xi_t(\vec{V}\cdot\hat{e}_t)\hat{e}_t - \xi_n(\vec{V}\cdot\hat{e}_n)\hat{e}_n \qquad [S5]$$

Using the small amplitude approximation, the propulsive force generated by the sperm flagellum at a given $x$ and $t$ is described by Eq. S6:

$$f_x \approx (\xi_n - \xi_t)\frac{\partial y}{\partial t}\frac{\partial y}{\partial x} \qquad [S6]$$

The total force in the horizontal direction ($\hat{x}$) must be calculated by integrating the force over the entire flagellum in one period (Eq. S7).

$$\langle f_x \rangle \approx \frac{1}{LT}\int_0^T dt \int_0^L dx\, (\xi_n - \xi_t)\frac{\partial y}{\partial t}\frac{\partial y}{\partial x} \qquad [S7]$$

By assuming that for $i > 1$, $y_i \ll y_1$, then Eq. S7 yields to Eq. S8

$$\langle f_x \rangle \approx -\frac{1}{2}(\xi_n - \xi_t)y_1^2\omega_1 k \qquad [S8]$$

And consequently the propulsion velocity can be described by Eq. S9. As can be seen, the propulsion velocity is correlated to the amplitude, frequency, and vector number of the sine wave with the greatest amplitude.

$$v_{propulsion} \approx -\frac{1}{2}\left(\frac{\xi_n}{\xi_t} - 1\right) y_1^2 \omega_1 k \qquad [S9]$$

In addition to the velocity generated in the $\hat{x}$ direction, the general force generated in the $\hat{y}$ direction for a segment of the flagellum is described by Eq. S10.

$$f_y \approx -\xi_n \frac{\partial y}{\partial t} + (\xi_n - \xi_t) \frac{\partial y}{\partial t}\left(\frac{\partial y}{\partial x}\right)^2 \qquad [S10]$$

Likewise, the general force in the $\hat{y}$ direction is the integral of $f_y$ over a period and the whole flagellum. Obviously, the first term of Eq. S10 vanishes after integration over the whole flagellum. Therefore, the generated force in the $\hat{y}$ direction is

$$\langle f_y \rangle \approx \frac{(\xi_n - \xi_t)}{LT} \int_0^T dt \int_0^L dx \left(-\sum y_i \omega_i \cos(kx - \omega_i t + \phi_i)\right)\left(\sum y_i k \cos(kx - \omega_i t + \phi_i)\right)^2. \qquad [S11]$$

By assuming that the sperm flagellum is composed of two sine waves, Eq. S11 reduces to

$$\langle f_y \rangle \approx \frac{(\xi_n - \xi_t)}{LT} \int_0^T dt \int_0^L dx(-y_1^2 y_n k^2 \cos(kx - \omega_n t + \phi_2) \cos^2(kx - \omega_1 t + \phi_1)(\omega_n + 2\omega_1)$$
$$+ -y_1 y_n^2 k^2 \cos(kx - \omega_1 t + \phi_1) \cos^2(kx - \omega_n t + \phi_2)(2\omega_n + \omega_1)). \qquad [S12]$$

Therefore, by assuming that $y_1 \gg y_n$, Eq. S12 yields to

$$\langle f_y \rangle \approx -y_1^2 y_n k^2 (\omega_n + 2\omega_1)(\xi_n - \xi_t)\frac{1}{LT}\left[\frac{\cos(2\phi_1 - \phi_n)}{4k(2\omega_1 - \omega_n)} + \frac{\cos(\phi_n)}{2\omega_n k} + \frac{\cos(2\phi_1 + \phi_n)}{12k(2\omega_1 + \omega_n)}\right]. \qquad [S13]$$

In particular, if $\omega_n = n\omega_1$, $\phi_1 = 0$, and $\phi_n = \phi$, then Eq. S13 reduces to

$$\langle f_y \rangle \propto -y_1^2 y_2 k(\omega_n + 2\omega_1)(\xi_n - \xi_t)\cos(\phi). \quad [S14]$$

Eventually, by considering that $v_{rotation} = \frac{\langle f_y \rangle}{\xi_n}$ and $\Omega_{intrinsic} = \frac{v_{rotation}}{L}$, the $\Omega_{rotation}$ is[4]

$$\Omega_{IN} \propto -y_1^2 y_n \frac{\omega_n + 2\omega_1}{L^3}\left(1 - \frac{\xi_t}{\xi_n}\right)\cos(\phi). \quad [S15]$$

The final equation obtained for the intrinsic angular velocity shows that depending on $\phi$, sperm movement can intrinsically have angular velocity, and therefore when the flow of the sperm medium is zero, the sperm trajectory can feature intrinsic curvature.

In the presence of fluid flow, as we clarified in the main text in Eq. 4, the angular velocity of the sperm is the sum of its intrinsic angular velocity and its rheotactic behavior as a response to external fluid flow[5]. According to Eq. S15, the intrinsic angular velocity of sperm can be either constructive ($\phi = 0 \rightarrow \Omega_{IN} = -\Omega_{max}$) or destructive ($\phi = \pi \rightarrow \Omega_{IN} = \Omega_{max}$) to the angular velocity imposed by the fluid flow. The impact of the intrinsic angular velocity on the trajectory of sperm is demonstrated in Fig. 1(g), which demonstrates the constructive effect of the intrinsic curvature leads to a decrease in the withdrawal distance. Moreover, the destructive effect of the intrinsic angular velocity is shown as well, leading to an increase in the withdrawal distance. As a result, the withdrawal distance can vary slightly depending on the $\phi$. To perform these simulations, it was assumed that the sperm oscillation frequency is constant, and consequently by plugging Eq. S9 into Eq. S15, a linear relation between intrinsic angular velocity and propulsion velocity (Eq. S16) can be obtained:

$$\Omega_{IN} \propto v_{propulsion}\left(\frac{\xi_t}{\xi_n}\right)\frac{y_n}{L^2}\cos(\phi) \quad [S16]$$

Two important results from these calculations include: (1) the withdrawal distance of the sperm cell is not significantly influenced by the intrinsic angular velocity of the sperm, and the major part of the sperm rotational movement is due to sperm rheotaxis; (2) even if the intrinsic curvature is not assumed to be negligible, as is demonstrated in Fig. 1(g) and Eq. S16, the sperm with higher velocity have higher corresponding intrinsic rotations. This means that the intrinsic angular velocity maintains the sperm with higher velocities closer to each other before the stricture. That is, for two highly motile sperm, the sperm with higher velocity and a destructive intrinsic rotation and the sperm with lower velocity and the constructive intrinsic rotation are not distinct. Therefore, the intrinsic rotation of the sperm is consistent with the fierce competition phenomenon among highly motile sperm.

**Experimental measurement of intrinsic angular velocity**

To experimentally confirm the existence of the intrinsic angular velocity, we observed the sperm swimming when the flow of the sperm medium was zero (Movie S1). The sperm trajectory in this zone is depicted in Fig. S2(a). Interestingly, most of the sperm (90%) in the zone with zero medium flow were rotating clockwise, and as can be seen in trajectories extracted for four different sperm (Fig. S2(b)), the curvature of the sperm remained roughly constant over time. We measured the intrinsic angular velocity of 28 sperm with different velocities (Fig. S3) and determined the mean value for their angular velocity was $0.12 \pm 0.06 \ s^{-1}$.

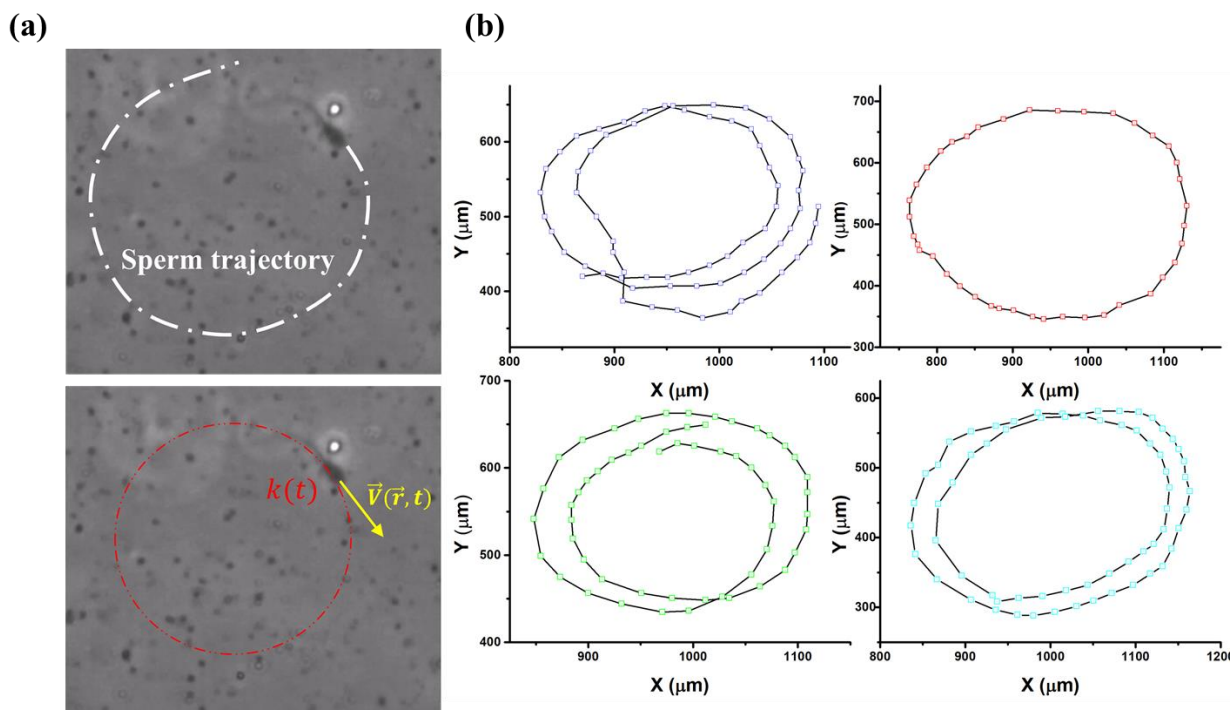

**Fig. S2 Sperm intrinsic angular velocity and curvature.** (a) Sperm pseudo-circular trajectory with a curvature of $k(t)$, which is roughly constant over time. (b) Trajectories of four different sperm in intrinsic rotation mode. The corresponding curvature of the trajectories were $8.24 \pm 1.24 \times 10^{-3} \mu m^{-1}$ (top-left), $12.4 \pm 2.32 \times 10^{-3} \mu m^{-1}$ (top-right), $8.08 \pm 0.94 \times 10^{-3} \mu m^{-1}$ (bottom-left), and $9.41 \pm 1.06 \times 10^{-3} \mu m^{-1}$ (bottom-right)

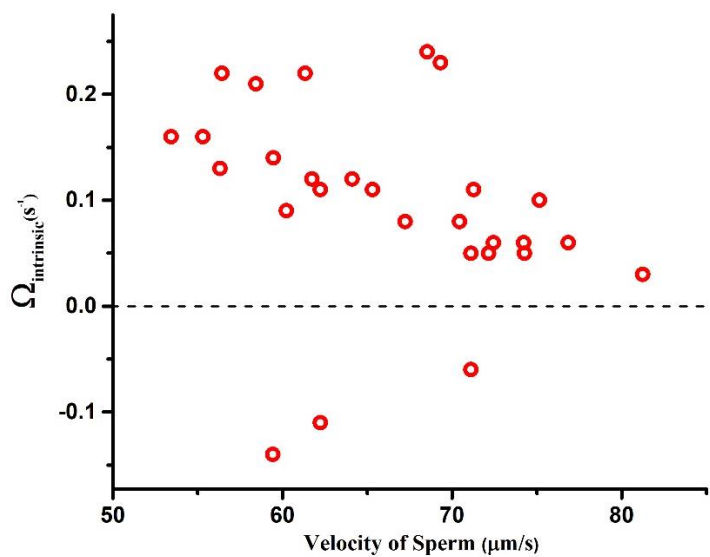

**Fig. S3 Intrinsic angular velocities measured for sperm when the external flow was zero.** Most of the sperm (~90%) swam in a clockwise direction ($\Omega_{IN} > 0$).

**Impact of stricture mouth angle on sperm motion**

The characteristics of the sperm butterfly-shaped motion highly depends on the stricture mouth angle, β. To demonstrate the impact of β on the sperm motion, the equations of motion were solved for different angles and the results are shown in Fig. S4. As can be seen, for β ~180º (Fig. S4(a)), the withdrawal distances of the sperm are very large, and therefore we can conclude that the CI of sperm with velocities between 40–80 µm/s is very low, and consequently the sperm cannot accumulate nearby the stricture. In addition, when sperm reach the other sidewall, the shear rate at the contact point in the $\hat{n}$ direction is inadequate (0.06–0.054 $s^{-1}$) to rotate sperm upstream, causing the sperm to move towards the downstream direction. When we decrease β to 130º (Fig. S4(b)), the CI starts to increase and accumulation of sperm close to the stricture begins to occur. Sperm with velocities in the range of ~70–80 µm/s can rotate upstream, and therefore these sperm are the only ones with a chance of passing through the stricture. For a β of 40º (Fig. S4(c)), the CIs are very low in comparison with β = 80° (Fig. 1 in the main text), and accordingly the chance of sperm to pass through the junction is very low. Moreover, only sperm with velocities in the range of $65 - 80$ µm/s are able to get closer than 5 µm to the stricture (this value is established as the proximity zone in the main text). As the angle decreases to 10º (Fig. S4(d)), the CIs are so low that none of the sperm with velocities in the range of 40–80 µm/s are able to enter the proximity zone of the stricture. Unlike the large angles, in this case the shear rate (5.64–6.12 $s^{-1}$) at the contact points can reorient sperm towards the upstream direction.

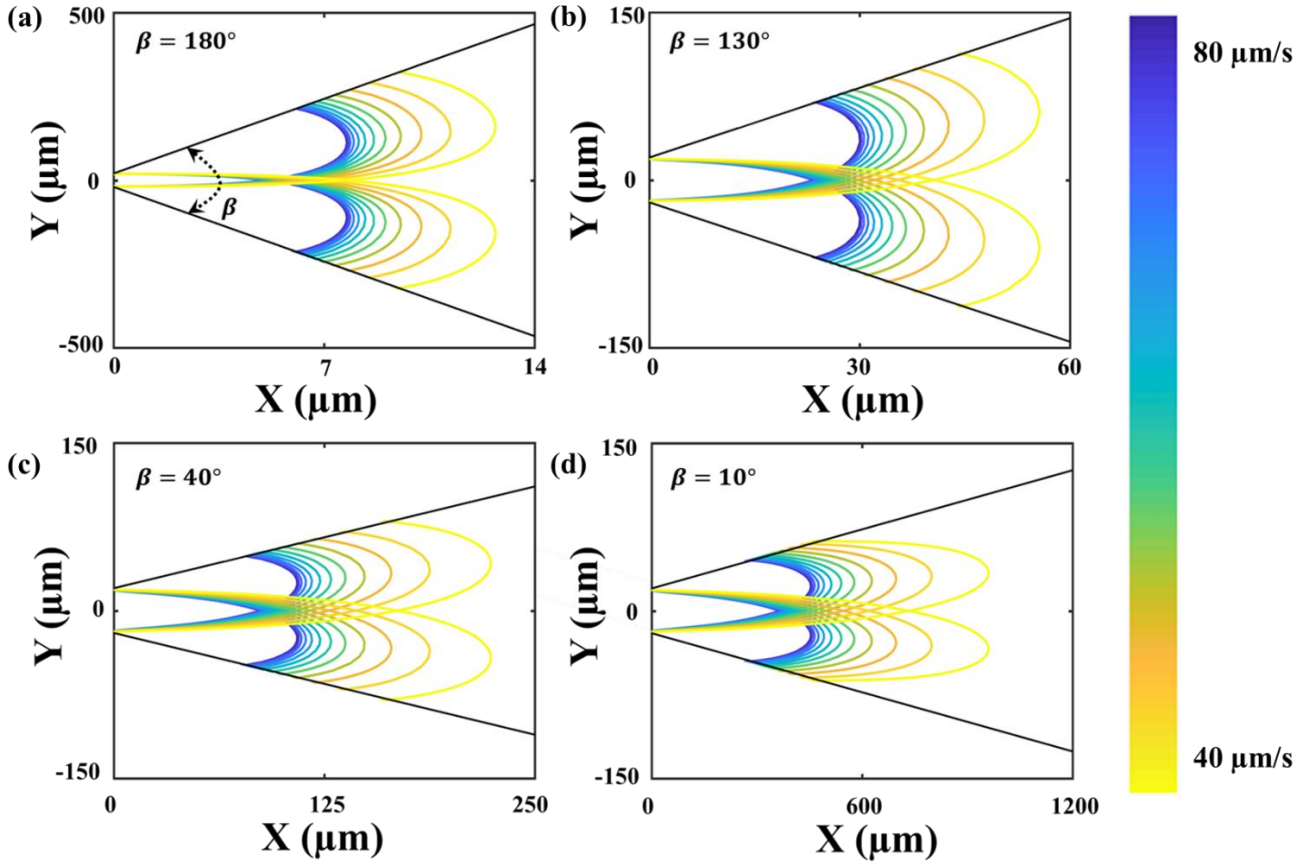

**Fig. S4 Impact of stricture mouth angle on the butterfly-shaped motion of sperm.** (a) An extremely wide stricture angle (β ~180º) leads to long withdrawal distances and thus low CIs. The low shear rates at the contact points are inadequate to reorient sperm upstream in the stricture direction. (b) Once β is 130º, a few sperm with velocities in the range of ~70–80 µm/s reorient upstream, while the rest of the sperm cannot return to the proximity of the stricture. (c) In acute stricture openings, the shear rate is enough to reorient sperm upstream and the butterfly-shaped motion occurs. However, for β = 40º, only a few sperm with velocities in the range of ~ 65–80 µm/s can get closer than 5 µm to the stricture. (d) At extremely acute angles (β = 10º), long withdrawal distances result. In addition, the proximity condition is satisfied for none of the sperm.

**Fokker-Planck equation**

The Fokker-Planck equation (Eq. S17) describes the evolution of the probability density function over time[7],

$$\frac{\partial P(x,t)}{\partial t} = \frac{\partial}{\partial x}\left(\frac{\partial x}{\partial t}P(x,t) - D_x\left(\frac{\partial P(x,t)}{\partial x}\right)\right) \quad [S17]$$

in which $P(x,t)dx$ is the likelihood of the sperm location to be between $x$ and $x + dx$, and $D_x$ is the translational diffusion coefficient of the sperm. For large Péclet numbers ($Pe = \frac{\tau_D}{\tau_v} \gg 1$) and the steady state condition ($\frac{\partial P(x,t)}{\partial t} = 0$), Eq. S17 reduces to

$$P(x,t)dx = Cdt. \quad [S18]$$

Considering the normalization $\int_0^L dx P(x,t) = \int_0^T dt C = 1$, Eq. S19 leads to the probability of the sperm being between $x$ and $x + dx$

$$P(x,t)dx = \frac{dt}{T} \quad [S19]$$

in which $dt$ is the time elapsed for sperm to swim the distance of $dx$. Therefore, the probability of the sperm to be closer than $a$ to the stricture is

$$pr\{X \leq a\} = \int_0^a dx P(x,t) = \int_0^{T_a} \frac{dt}{T} + \int_{T'_a}^T \frac{dt}{T} = \frac{T_a + T - T'_a}{T}. \quad [S20]$$

in which $T_a$ and $T'_a$ are the corresponding times of $x = a$, in which because of the periodic motion, the sperm pass through this situation twice.

**Butterfly-shaped motion in human sperm**

The similarity between human and bovine sperm in terms of the shape and swimming mechanism suggests that the motion of human sperm before the stricture is like that of bovine sperm. To experimentally confirm this similarity, we experimentally observed human sperm motion before the stricture (Movie S3). As was expected, we observed the butterfly-shaped swimming path (Fig. S5(a)). The withdrawal distances were measured and reported in Fig. S5(b). We also observed the human sperm motion in transfer, rotation, and boundary swimming modes (Fig. S6(a)). Fig. S6(b) presents the time elapsed in these modes.

The data extracted for the human sperm (Fig. S5 and S6) confirms the similarity in the data trends between the human and bovine sperm. However, the slower swimming speed of human sperm led to longer periods and withdrawal distances. Moreover, for very slow sperm ($v < 35$ µm/s) the time required for the transfer mode was longer than the time required for a sperm to completely reorient itself upstream, and thus the reorientation happens before reaching the other sidewall, preventing transfer to the other sidewall. This leads to slow sperm appearing to be static in the observer frame. In fact, once their swimming direction is completely aligned with the flow streamlines, their propulsive force is neutralized by the fluid flow, as can be seen in Movie S3.

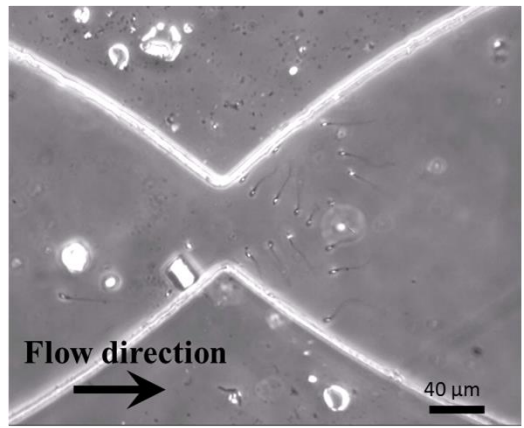 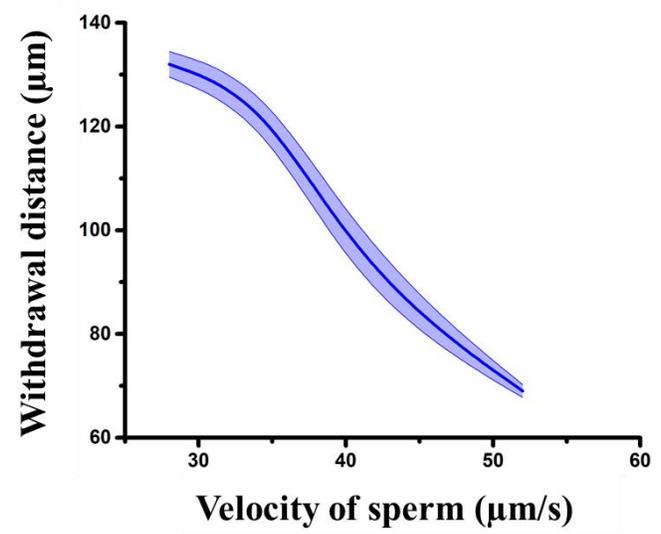

**Fig. S5 The butterfly-shaped motion of sperm.** (a) Human sperm swim on a butterfly-shaped path and the total swimming direction is counter to the flow. (b) The withdrawal distance was extracted for different sperm with different velocities.

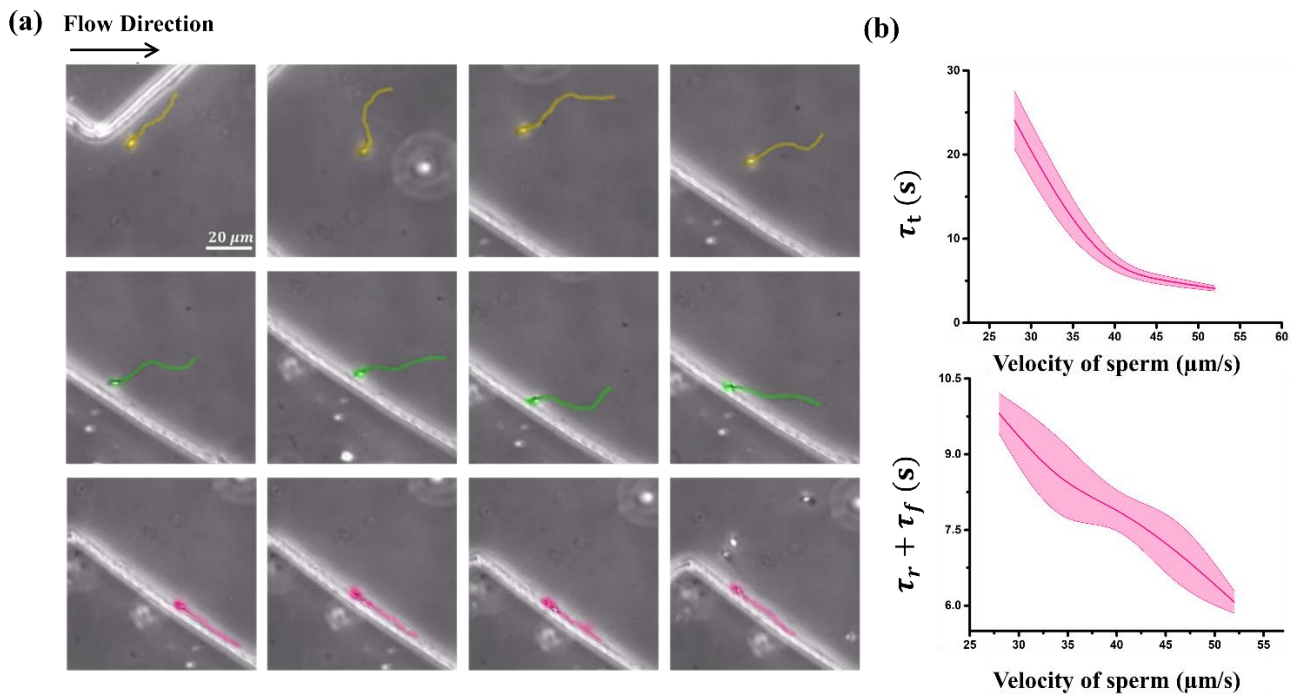

**Fig. S6 Transfer, rotation, and boundary swimming modes with corresponding times.** (a) The sperm swimming in the transfer, rotation, and boundary swimming modes are illustrated in the first, second, and third rows, respectively. To visualize the sperm in each frame, three colors are used. The scale bar and flow direction are the same for all pictures. (b) The time elapsed in the transfer ($\tau_t$) and combined rotation and boundary swimming modes ($\tau_r + \tau_f$).

**Threshold sperm velocity**

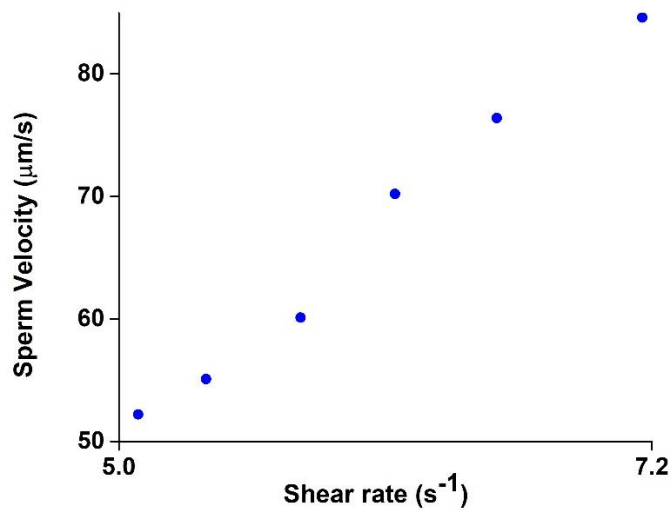

**Fig. S7 Threshold sperm velocity for different shear rates of the stricture.**